\documentclass[twocolumn,prb,superscriptaddress,showpacs,citeautoscript,floatfix]{revtex4}

\usepackage{amsfonts}
\usepackage{amssymb}
\usepackage{amsmath}
\usepackage{graphicx}
\usepackage{dcolumn}   
\usepackage{bm}        
\usepackage{flafter}
\usepackage{hyperref}

\renewcommand{\vec}[1]{\ensuremath{\boldsymbol{#1}}}

\begin{document}


\title{Effect of pressure on the Raman modes of antimony}

\author{X. Wang}
\affiliation{Max-Planck-Institut f\"{u}r Festk\"{o}rperforschung,
Heisenbergstrasse 1, D-70569 Stuttgart, Germany}

\author{K. Kunc}
\altaffiliation[Permanent address: ]{Institut des Nanosciences de
Paris, CNRS and Universit\'e Pierre and Marie Curie,
     140 rue de Lourmel, F-75015 Paris, France}
\affiliation{Max-Planck-Institut f\"{u}r Festk\"{o}rperforschung,
Heisenbergstrasse 1, D-70569 Stuttgart, Germany}

\author{I. Loa}
\affiliation{Max-Planck-Institut f\"{u}r Festk\"{o}rperforschung,
Heisenbergstrasse 1, D-70569 Stuttgart, Germany}

\author{U. Schwarz}
\affiliation{Max-Planck-Institut f\"{u}r Chemische Physik
fester Stoffe, N\"{o}tnitzer Strasse 40, D-01187 Dresden,
Germany}

\author{K. Syassen}
\email[E-mail:~]{k.syassen@fkf.mpg.de}
\affiliation{Max-Planck-Institut f\"{u}r Festk\"{o}rperforschung,
Heisenbergstrasse 1, D-70569 Stuttgart, Germany}

\date{28 August 2006}

\begin{abstract}

The effect of pressure on the zone-center optical phonon modes of
antimony in the $A7$ structure has been investigated by Raman
spectroscopy. The A$_g$ and E$_g$ frequencies exhibit a pronounced
softening with increasing pressure, the effect being related to a
gradual suppression of the Peierls-like distortion of the $A7$ phase
relative to a cubic primitive lattice. Also, both Raman modes broaden
significantly under pressure. Spectra taken at low temperature indicate
that the broadening is at least partly caused by phonon-phonon
interactions. We also report results of \textit{ab initio}
frozen-phonon calculations of the A$_g$ and E$_g$ mode frequencies.
Presence of strong anharmonicity is clearly apparent in calculated
total energy versus atom displacement relations. Pronounced
nonlinearities in the force versus displacement relations are observed.
Structural instabilities of the Sb-A7 phase are briefly addressed in
the Appendix.

\end{abstract}

\bigskip

\pacs{PACS:
78.30.-j  
63.20.-e  
71.15.Nc  
62.50.+p  
}

\maketitle

\section{Introduction}

Like other semi-metals of the group V-A, antimony crystallizes at
ambient conditions in the trigonal $A7$ structure (space group
$R\overline{3}m$, No. 166) \cite{Donohue}. Antimony is known to
transform at $\approx$8.5~GPa to a high-pressure phase \cite{Schif81}
that is referred to as Sb-II in the recent literature. At 28~GPa Sb
converts to the body-centered cubic modification Sb-III
\cite{Aoki83,Iwasa97}. Diffraction studies \cite{McMah00,Schwa03} show
Sb-II to adopt a tetragonal self-hosting structure with an
incommensurate guest-host arrangement. In addition, incommensurate
modulations occur in Sb-II \cite{Schwa03}. The structural sequence for
Sb under pressure is further complicated by the observation of another
incommensurate phase, called Sb-IV, in a narrow pressure range from 8
to 9~GPa (upstroke), i.e. intermediate between Sb-I and Sb-II
\cite{Degty04a,Degty04}.

\begin{figure}[b]\centering
\includegraphics[width=50mm,clip]{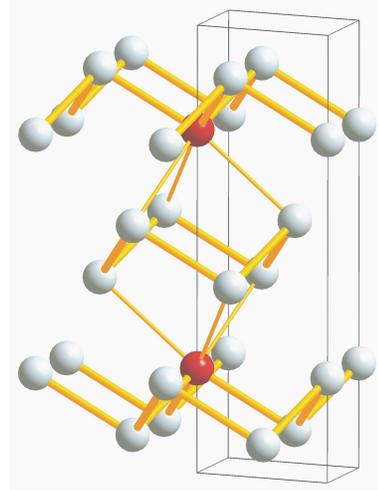}
\caption{\label{fig1} (Color online) The $A7$ structure of Sb. Two
distortion modes lead from the parent cubic-primitive (cP1) to the A7
(hR2) structure: a rhombohedral elongation along a cubic [111]
direction and a pairing of the original cubic (111) lattice planes.
Atoms in red mark opposite corners of the distorted cube. The resulting
layered structure has a 3+3 atomic coordination.  The non-primitive
hexagonal unit cell is indicated in the figure. }
\end{figure}

Here we are interested mainly in the effect of pressure on optical
phonon modes of the $A7$ phase of Sb. The $A7$ structure
(Fig.~\ref{fig1}) is a rhombohedrally distorted variant of a cubic
primitive structure. The formation of paired layers results in a
doubling of the number of atoms in the primitive rhombohedral cell. In
this context, the term Peierls distortion is appropriate because the
pairing is driven by a gain in electronic energy (formation of a
pseudo-gap in the electronic density of states at the Fermi level
$E_{\rm F}$) which outweighs the cost in Ewald energy \cite{Chang86}.
In the following we will mostly refer to the hexagonal description of
the structure. The non-primitive hexagonal unit cell contains six
atoms. For conversion relations between rhombohedral and hexagonal
cells we may refer to Ref.\mbox{~}\onlinecite{Sharp71}
or\mbox{~}\onlinecite{Needs86}.

Group theory for the $A7$ structure predicts three zone-center optical
modes, i.e. a singly degenerate A$_{\rm 1g}$ mode (totally symmetric)
and a doubly degenerate E$_{\rm g}$ mode. The atoms are displaced along
the $C_3$ axis in the A$_{\rm 1g}$ mode and perpendicular to that axis
in the E$_{\rm g}$ mode. The A$_{\rm 1g}$ mode is the 'Peierls
distortion mode'. In the unfolded Brillouin zone of the cubic primitive
structure, the A$_{\rm 1g}$ mode corresponds to a longitudinal acoustic
mode at the R-point, whereas the E$_{\rm g}$ mode corresponds to a
transverse acoustic mode at R.

Previous Raman studies of antimony at ambient pressure have addressed
the resonance effects \cite{Renucci73}, the second-order scattering by
optical modes \cite{Lannin75}, the effect of temperature on the
first-order modes \cite{Hoehne77},  and the coupling of the first-order
Raman modes to electronic excitations \cite{Bansa86}. The coherent
A$_{\rm 1g}$ phonon generation in Sb by short laser pulses is the
subject of several more recent reports \cite{Cheng90}. Raman
experiments on Sb under pressure were performed by Richter \textit{et
al.} \cite{Richt} who measured frequency shifts up to 0.6~GPa and
observed a softening of the zone-center optical modes, not only for Sb,
but also the $A7$ phases of As and Bi. Optical phonons in As were
studied up to 30 GPa using Raman spectroscopy \cite{Beist90a} whereas
in the case of Bi femtosecond pump and probe techniques were applied to
explore the effect of pressure up to $~$3 GPa on the modulation of the
optical response by coherent A$_{\rm 1g}$ phonons
\cite{Kasami04,Kasami06}.

In this work, we have investigated the full stability range of the $A7$
phase of Sb using Raman spectroscopy. We also report Raman spectra of
Sb-II measured up to 16~GPa. Our discussion focuses on results for the
$A7$ phase, specifically the pressure-induced softening of mode
frequencies up to the first phase transition and the observed strong
pressure dependence of the Raman line widths. Experimental results for
phonon frequencies of the $A7$ phase are compared to frozen-phonon
calculations performed within the density functional framework using
optimized structural parameters. A side aspect of our calculations
concerning structural instabilities of the A7 phase in the vicinity of
the Sb-I to Sb-II transition is presented in the Appendix.

\begin{figure}[b]\centering
\includegraphics[width=70mm,clip]{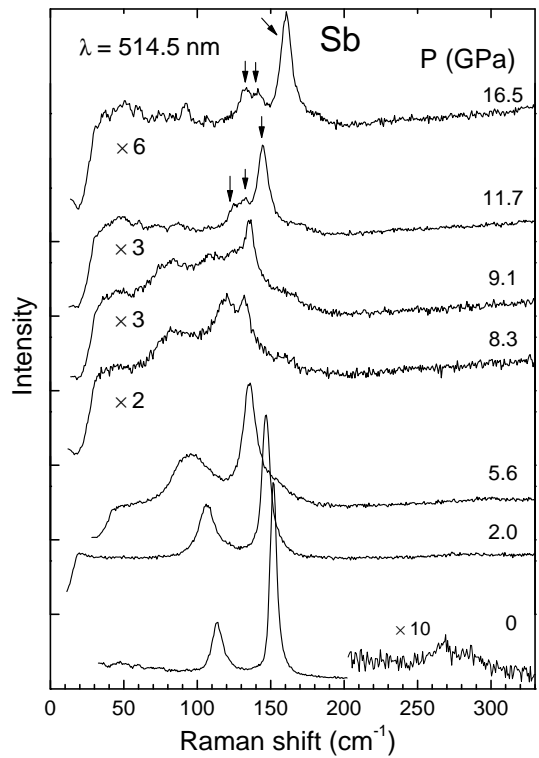}
\caption{\label{fig2} Raman spectra of antimony at different pressures
($T = 300$~K). Spectra are scaled to about the same maximum intensity;
the scaling factors are given in the figure. The broad feature seen at
about 270~cm$^{-1}$ (see blown-up section of the bottom spectrum) is
attributed to second-order Raman scattering of the $A7$ phase. The
spectra at 8.3~GPa and 9.1~GPa correspond to a phase mixture of $A7$
and high pressure phase(s), those at 11.7 and 16.5 GPa to Sb-II. Arrows
mark Raman features characteristic of Sb-II.}
\end{figure}

\section{Raman spectroscopy}

The Raman scattering measurements were performed on crystalline
antimony of 99.999\% purity. A diamond anvil cell (DAC) was used for
pressure generation. Experiments were initially performed with nitrogen
as a pressure medium and then, in order to avoid the Raman scattering
of solid nitrogen \cite{Schne92}, continued with the standard 4:1
methanol-ethanol mixture (M/E) as well as with helium; the latter
ensures the best hydrostatic conditions in combination with low
background. In all experiments, pressures were measured by the ruby
luminescence method \cite{Pierm75,Mao86}. Raman spectra as a function
of pressure were recorded in backscattering geometry (normal of the
sample surface parallel to the $C_3$ axis) using a triple-grating
spectrometer (Jobin-Yvon T64000) in combination with a
liquid-nitrogen-cooled charge-coupled device (CCD) detector. The
resolution was set to $< 2$~cm$^{-1}$. An argon ion laser was used for
excitation at a wavelength of 514.5 nm. The laser was focused to a spot
of 50 $\mu$m in diameter. The power incident at the sample surface was
15-20 mW. At that power level we did not notice any indication for
laser-induced heating of the sample when mounted inside the DAC.

\begin{figure}[tb]\centering
\includegraphics[width=75mm,clip]{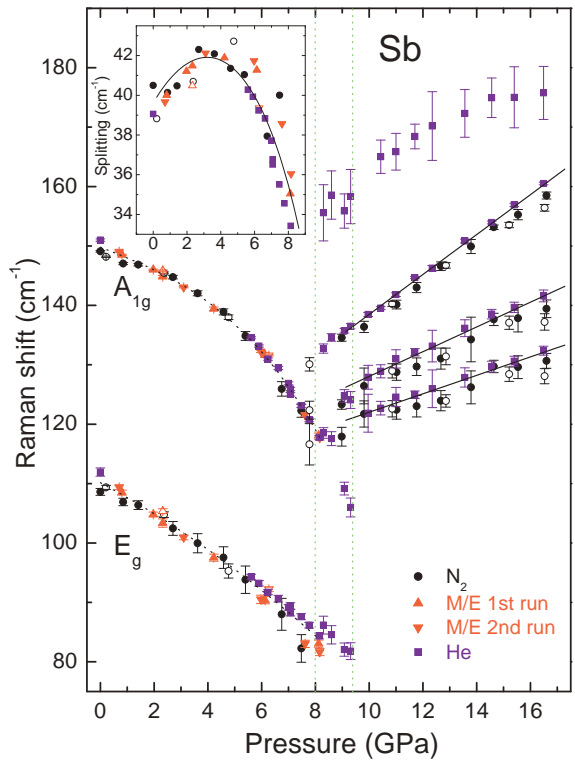}
\caption{\label{fig3} (Color online) Raman peak frequencies of Sb as a
function of pressure at room temperature. Closed symbols are for
increasing pressure, open symbols for decreasing pressure. Data were
collected in runs using different pressure media. The transition regime
between the $A7$ phase and phase II is indicated by vertical dashed
lines. Lines passing through the data points represent results of least
squares fits (for parameters see Table~\ref{tab1}). The inset shows the
splitting of the optical mode frequencies of the $A7$ phase as a
function of pressure. There, the line is a guide to the eye.}
\end{figure}

Figure~\ref{fig2} shows a subset of Raman spectra of antimony measured
at room temperature and at different pressures up to 16.5 GPa. Spectra
at 0, 2, and 5.6~GPa are for the A7 phase, those at 8.3 and 9.1~GPa are
for phase mixtures, and those above 10 GPa are for Sb-II. The
frequencies of the two observed Raman modes of the $A7$ phase decrease
with increasing pressure. The A$_{\rm 1g}$ feature is slightly
asymmetric at low pressure, and with increasing pressure a shoulder
develops at its high energy side. A broad feature between 235 and
320~cm$^{-1}$ seen at ambient pressure can be attributed to
second-order scattering by optical phonon modes \cite{Lannin75}. Under
applied pressure, that feature could not be separated well enough from
the background scattering to allow for useful information to be
obtained.

\begin{table}[tb]\centering
\caption{\label{tab1} Frequencies of the Raman features of the Sb-A7
and Sb-II phases. The pressure dependence is described by $\omega(P) =
\omega_0 + a\,P + b\,P^2 + c\,P^3$ where frequency $\omega$ is
in~cm$^{-1}$ and pressure $P$ in~GPa. For the Sb-I phase, the
zero-pressure frequencies $\omega_0$, the linear pressure coefficients
$A$, the quadratic one ($B$) for the E$_{\rm g}$ mode, and the cubic
coefficient $C$ for the A$_{\rm 1g}$ mode are given. The zero-pressure
mode Gr\"{u}neisen parameters $\gamma_0$ are determined from the linear
coefficients $a$ and the reported bulk modulus $B_0=39.7$~GPa
\cite{Degty04}. For the Sb-II phase, we list Raman peak frequencies
$\omega$ at 9.1~GPa and the average linear coefficients $\overline{a}$
for the pressure range from 9 to 16~GPa. The average values
$\overline{\gamma}$ of the mode Gr\"{u}neisen parameter given for that
range are estimated assuming a bulk modulus value of 100~GPa.}

\medskip

\setlength{\extrarowheight}{2pt}

\begin{ruledtabular}
\begin{tabular}{lcccccc}
A7 &Mode &$\omega_0$ &$a$ &$b$  &$c$ &$\gamma_0$\\
%
%
\hline
     &E$_{\rm g}$  &110(1) &-2.4(6)& -0.10(7)& 0 &-0.58\\
    &A$_{\rm 1g}$ &149.8(10) &-1.8(3) & 0 & -0.032(3) &-0.48\\
\hline
Sb-II &Mode & $\omega$(9.1 GPa)  &$\overline{a}$& & & $\overline{\gamma}$\\
%
%
\hline
     &1 &121(2) & 1.6(1) &&  &1.3\\
    &2   &126(2) &2.1(1) &&  &1.65 \\
    &3 &  135.6(5)  &3.36(4) &&  &2.45\\
%
\end{tabular}
\end{ruledtabular}
\end{table}

Schwarz \textit{et al.} \cite{Schwa03} pointed out that the transition
to Sb-II is sluggish and that a pure phase Sb-II can be obtained only
above $~$13(1)~GPa. On the other hand, Degtyareva \textit{et al.}
obtained a pure phase Sb-II at 9.1(1) GPa (upstroke). Our Raman spectra
from different runs indicate a transition regime extending from about 8
to 10~GPa.

After passing the transition regime, the Raman spectra look quite
different from those of the $A7$ phase. At least three new Raman
features appear as indicated by the arrows in Fig.~\ref{fig2}. All of
them shift to higher frequency with increasing pressure. We could not
reproducibly identify other distinct Raman features of Sb-II, except
for a broad shoulder on the high energy side of the most intense Raman
peak.

It has been reported that, at ambient pressure, the E$_{\rm g}$ and
A$_{\rm 1g}$ Raman features exhibits a Fano-like line shape which is
believed to arise from the interaction between electronic intraband
scattering and one-phonon scattering \cite{Bansa86}. These asymmetries
are weak, however. So, for the purpose of extracting peak positions and
approximate line width parameters from the Raman data for the A7 phase,
a fitting by symmetric pseudo-Voigt line shapes was found to be
adequate. For Sb-II, peak frequencies were determined by simply
locating intensity maxima. The obtained frequencies of Raman features
of Sb for pressures up to 16~GPa are shown in Fig.~\ref{fig3}. Within
the scatter of the frequency data, the pressure effects are fully
reversible.

The observed ambient-pressure A$_{\rm 1g}$ and E$_{\rm g}$ mode
frequencies are 149.8~cm$^{-1}$ and 110~cm$^{-1}$ at room temperature
[values measured for one of the samples at 2~K are 155.7(7) and
117.0(3)~cm$^{-1}$]. For a pressure of 7.5~GPa, the total relative
shift is -18\% and -24\% for the A$_{\rm 1g}$ and E$_{\rm g}$ modes.
The magnitude of these shifts is similar to what has been found for the
optical modes of arsenic in the more extended stability range (0 -- 25
GPa) of its A7 phase \cite{Beist90a}. The nonlinear $\omega(P)$
behavior of the E$_{\rm g}$ mode of Sb can be described by the usual
combination of a linear and quadratic pressure coefficient, that of the
A$_{\rm 1g}$ mode is approximated well by a combination of a linear and
a cubic term in pressure. Sets of polynomial parameters characterizing
the pressure dependence of mode frequencies and mode Gr\"{u}neisen
parameters are given in Table~\ref{tab1}.

The frequencies of the Raman bands attributed to the phase Sb-II are
121, 126, and 135.6~cm$^{-1}$ at 9.1~GPa. The frequencies of all three
features exhibit an essentially linear pressure dependence up to 16
GPa. The corresponding mode Gr\"{u}neisen parameters $\overline{\gamma}$
given in Table~\ref{tab1} were estimated using a bulk modulus value of
$B_0 = 100$~GPa as extracted from pressure-volume data given in
Ref.\mbox{~}\onlinecite{Degty04a}.

The structure of Sb-II consists of an incommensurate guest-host
arrangement. For the host structure with tetragonal symmetry and space
group $I4/mcm$\mbox{ }\onlinecite{Degty04a}, a group theory analysis
gives five zone-center Raman-active modes of symmetry A$_{\rm 1g }$,
B$_{\rm 1g }$, B$_{\rm 2g}$, E$_{\rm g }$(1), and E$_{\rm g }$(2). The
corresponding displacement patterns are those of the aluminum
sublattice in CuAl$_2$ \cite{Grin06}. Only three Raman features could
be clearly identified in our spectra for Sb-II. Not having investigated
polarization effects, we do not propose an assignment at this point.
So, concerning Sb-II, the main result of the present study is that the
frequencies of Raman peaks presumably due to optical phonons are
similar to that of the A$_{\rm 1g}$ mode of Sb-I and, different from
Sb-I, the optical phonons of Sb-II harden with increasing pressure and
the values of the mode Gr\"{u}neisen parameters are larger than one.

\begin{figure}[tb]\centering
\includegraphics[width=70mm,clip]{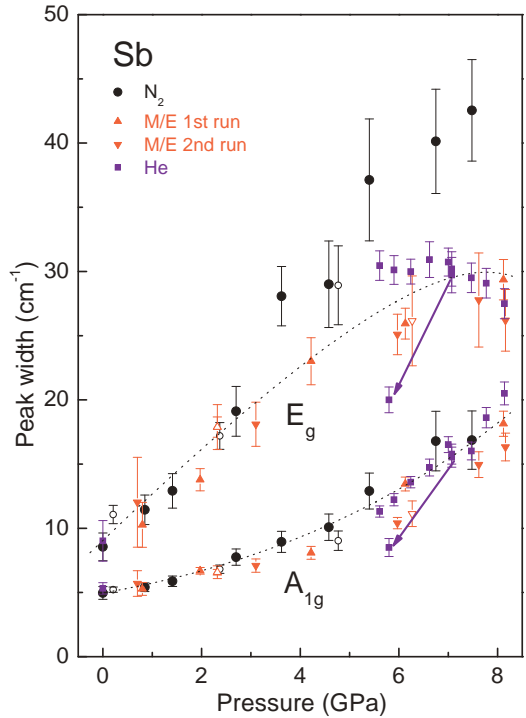}
\caption{\label{fig4} (Color online) Raman line widths (FWHM) of A7
antimony as a function of pressure at room temperature. Different
shapes of symbols are used for different pressure media. Closed symbols
are for increasing pressure, open symbols for decreasing pressure.
Dashed lines are guides to the eye. Note that the pressure-induced
broadening is fully reversible. Arrows indicate the change in width
upon cooling to 5 K (cf. spectra shown in Fig.~\ref{fig5}).}
\end{figure}

\begin{figure}[b]\centering
\includegraphics[width=70mm,clip]{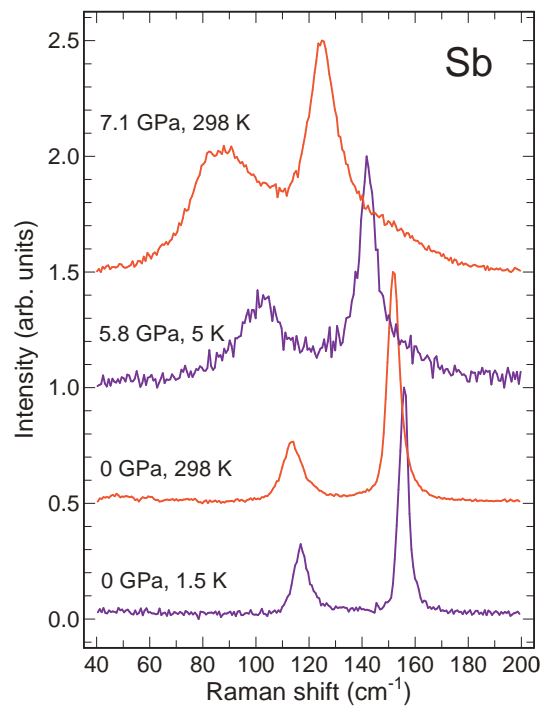}
\caption{\label{fig5} (Color online) A room-temperature Raman spectrum
of Sb at 7.1 GPa compared to the spectrum measured after cooling to 5
K. Upon cooling, the pressure dropped to 5.8 GPa. Spectra were measured
for a helium pressure medium.}
\end{figure}

In view of the observed changes in the phonon frequencies under
pressure, we briefly mention two related properties of Sb. First, there
is the negative slope of the melting line of Sb-I. The melting
temperature drops from 904 K at ambient to about 830 K at 5 to 5.7~GPa
(the triple point between liquid, Sb-I and high pressure phase)
\cite{Klement63,Khvos84}. Within a 'single-phase' approach to melting
and in the spirit of the Lindemann melting criterium, a microscopic
explanation of the negative melting slope would have to take into
account that at least part of the phonon spectrum softens considerably
with increasing pressure. The second remark concerns the
pressure-induced superconductivity of antimony. The superconducting
transition temperature $T_{\rm c}$ of Sb-I increases with increasing
pressure, reaching 0.7~K at 8.5~GPa \cite{Witti68}. At the phase
transition to Sb-II, $T_{\rm c}$ jumps to 3.7 K and then it drops
continuously to 2 K at 25 GPa \cite{Witti68}. Qualitatively, the
increase of $T_{\rm c}$ in the A7 phase can be understood to arise from
an increase of the density of states $N(E_{\rm F})$ (i.e. the filling
of the pseudogap, cf. Refs.\mbox{~}\onlinecite{Haeus02}
and\mbox{~}\onlinecite{Ormeci04}) and, in view of the strong phonon
softening, an enhanced electron-phonon coupling \cite{Chang86}. In the
phase Sb-II, $N(E_{\rm F})$ is much larger compared to that of Sb-I at
ambient pressure \cite{Haeus02,Ormeci04}. The effect of pressure on
$N(E_{\rm F})$ of Sb-II appears to be not very pronounced
\cite{Ormeci04}. So, one can speculate that a hardening of relevant
phonon modes, as indicated by our Raman data for optical modes,
accounts for the decrease of $T_{\rm c}$ with increasing pressure in
Sb-II.

Besides a large phonon softening in Sb-I, the second very pronounced
effect of pressure is the change in line widths of the A$_{\rm 1g}$ and
E$_{\rm g}$ Raman peaks. The pressure dependence of the peak widths
(FWHM) is shown in Fig.~\ref{fig4}. Error bars take into account some
ambiguity in separating a peak from the background and/or shoulders.
Between ambient pressure and 8 GPa, the width of the A$_{\rm 1g}$
feature increases by about a factor of three, from $\sim$6~cm$^{-1}$ to
$\sim$17~cm$^{-1}$. The E$_{\rm g}$-related peak starts with a width of
$\sim$9~cm$^{-1}$ at ambient and the width increases substantially to
$\sim$30~cm$^{-1}$ at 8~GPa. For the $N_2$ medium (solid at 2.5~GPa and
changing phase at 6 GPa), the width becomes even larger under pressure;
so, the observed line width of the E$_{\rm g}$ feature seems to depend
sensitively on non-hydrostaticity or shear strength of the pressure
medium. The data obtained with alcohol and helium medium are considered
to reflect the effect of truly hydrostatic stress on the Raman line
widths. It is important to note that the pressure-induced broadenings
were found to be fully reversible within experimental scatter. This
also applies to runs which, during upstroke, passed into the Sb-II
phase regime.

We have performed one high-pressure experiment at low temperature. The
sample was loaded in He medium, taken to 7.1 GPa at 300 K, and then
cooled to 5 K. During cooling the pressure dropped somewhat to 5.8 GPa
due to experimental factors. The high-pressure spectra at 300~K and 5~K
are shown in Fig.~\ref{fig5}, together with room- and low-temperature
spectra measured at ambient pressure. At zero pressure and when
cooling, the width of the A$_{\rm 1g}$ drops from 5.4 to 4.1 cm$^{-1}$,
that of the E$_{\rm g}$ mode from 8.9 to 6.6 cm$^{-1}$. The
low-temperature widths are larger compared to those reported in Ref.
\cite{Hoehne77}, presumably due to crystal imperfections induced by the
sample preparation procedures. At high pressure the change in width is
14 $\rightarrow$ 8.5 cm$^{-1}$ for A$_{\rm 1g}$ and 28 $\rightarrow$ 20
cm$^{-1}$ for E$_{\rm g}$. The latter changes are indicated by arrows
in Fig.~\ref{fig4}. These changes in peak widths demonstrate that about
half of the phonon decay rate at room temperature is thermally
activated. Spectral weight components like the shoulder on the
high-frequency side of A$_{\rm 1g}$ also depend on temperature.

A possible explanation of the line widths changes under pressure would
be the coupling to acoustic phonons via cubic anharmonicity
\cite{Hoehne77,Scott74}. At ambient pressure, the E$_{\rm g}$ mode
frequency represents the lower frequency limit of the optical phonon
branches, separated by about 10 cm$^{-1}$ from the acoustical mode
regime \cite{Sharp71}. Based on the experimental phonon dispersion
relations \cite{Sharp71}, one can expect a large two-phonon density of
states (DOS) for acoustical modes spanning the range from about 80 to
160 cm$^{-1}$. So, decay of the zone-center optical phonons into two
acoustical modes should be energetically possible. In view of the
optical phonon softening and related strong anharmonicity (see also
discussion below), an increase of the third-order coupling strength
under pressure may be the dominant effect causing the line broadening.
At this point we hesitate to elaborate further on this scenario,
because we don't know the effect of pressure on the two-phonon DOS and,
furthermore, pressure-induced changes in electronic structure should be
considered as well.

At ambient pressure, antimony is a semimetal with a negative indirect
gap of $\sim$ 0.1 eV. The free carriers (density $\sim$
10$^{20}$/cm$^3$) occupy states at Brillouin zone edges, forming
pockets with small effective masses \cite{Falicov66,Dress71,Liu95}. A
coupling between the optical phonon modes and low-energy electronic
excitations appears to be supported by a Raman study \cite{Bansa86}.
Now, the pseudogap of Sb is expected to change significantly (towards
filling up) under pressure, leading to a major change in the coupling
of phonons to electronic states near the Fermi level \cite{Chang86}.
Furthermore, the response in the visible spectral range may be affected
also, in particular the so-called E$_2$ transition near 2.5~eV\mbox{~}
\onlinecite{Cardo64} which at ambient pressure is in resonance with the
exciting laser energy used in this study. So, we cannot neglect effects
related to the penetration depth of the laser light \cite{Klein82}.
Furthermore, resonance conditions for Raman scattering may also change.

\begin{table}[tb]
\caption{\label{tab2} Structural parameters of Sb-I obtained from DFT
calculations using the LDA and GGA schemes. The values refer to the
respective calculated zero-pressure volumes. The lattice parameters are
for the non-primitive hexagonal cell, the volumes are for the primitive
unit cell (2 atoms). The calculated values of bulk modulus $B_0$ and
its pressure derivative $B_0^{\prime}$ were obtained by fitting a
third-order Birch equation of state to the calculated pressure-volume
data (the attached errors reflect parameter correlations). Related
experimental results are listed in the lower part of the table.}

\medskip

\setlength{\extrarowheight}{2pt}

\begin{ruledtabular}
\begin{tabular}{lllllll}
       & $a$ ({\AA}) & $c/a$ & $u$ & $V_0$ ({\AA}$^{3}$) & $B_0$ (GPa) & $B_0^{\prime}$\\
\hline
LDA    &4.3003 &2.5406 &0.2368 &58.32(1) &46.7(6) &4.9(3) \\
GGA    &4.3779 &2.6209 &0.2334 &63.48(3) &34.4(6) &5.4(2)\\
\hline
4 K  &4.3007\footnotemark[1] &2.6093\footnotemark[1] &0.23362\footnotemark[1] &59.918 & & \\
298 K  &4.3084\footnotemark[1] &2.6167\footnotemark[1] &0.23349\footnotemark[1] &60.41 &39.7\footnotemark[2] & 4 \footnotemark[2]\\

298 K  & & & & &45.6\footnotemark[3]& \\
298 K  & & & & &40.2\footnotemark[4] & \\
\end{tabular}
\end{ruledtabular}

\medskip

\footnotetext[1]{\ \ X-ray diffraction, Ref.\mbox{~}\onlinecite{Barre63}}%
\footnotetext[2]{\ \ X-ray diffraction, $B_0^{\prime}$ fixed, Ref.\mbox{~}\onlinecite{Degty04}}%
\footnotetext[3]{\ \ Ultrasonic wave velocities, Ref.\mbox{~}\onlinecite{Epste65}, adiabatic value}%
\footnotetext[4]{\ \ Ultrasonic wave velocities, Ref.\mbox{~}\onlinecite{Hearm46}, adiabatic value}%

\end{table}

\begin{figure*}[tb]
\begin{minipage}[b]{0.49\hsize}\vspace{0pt}
\centerline{\includegraphics[height=94mm,clip]{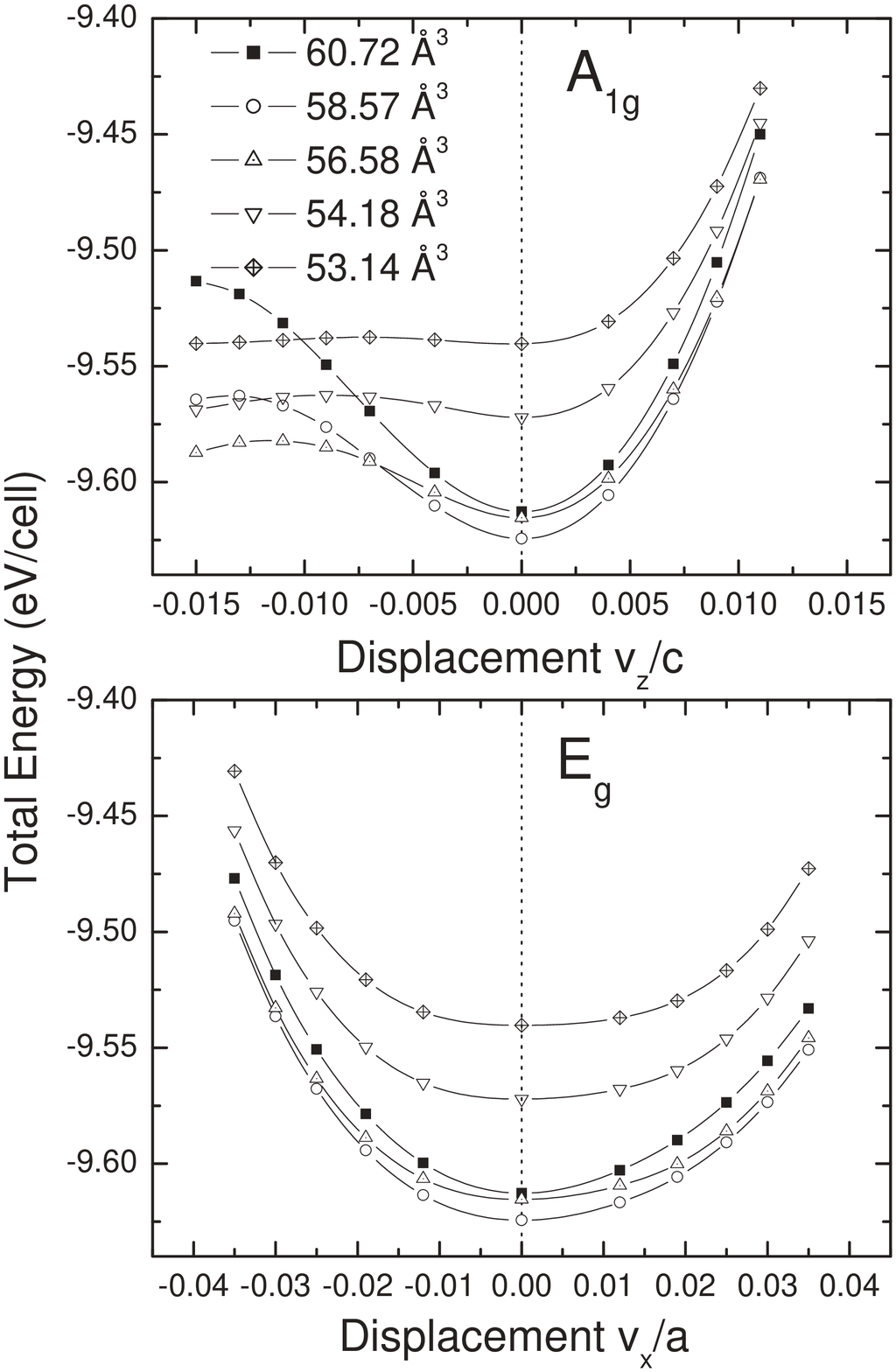}}
\end{minipage}
\hfill
\begin{minipage}[b]{0.49\hsize}\vspace{0pt}
\centerline{\includegraphics[height=93mm,clip]{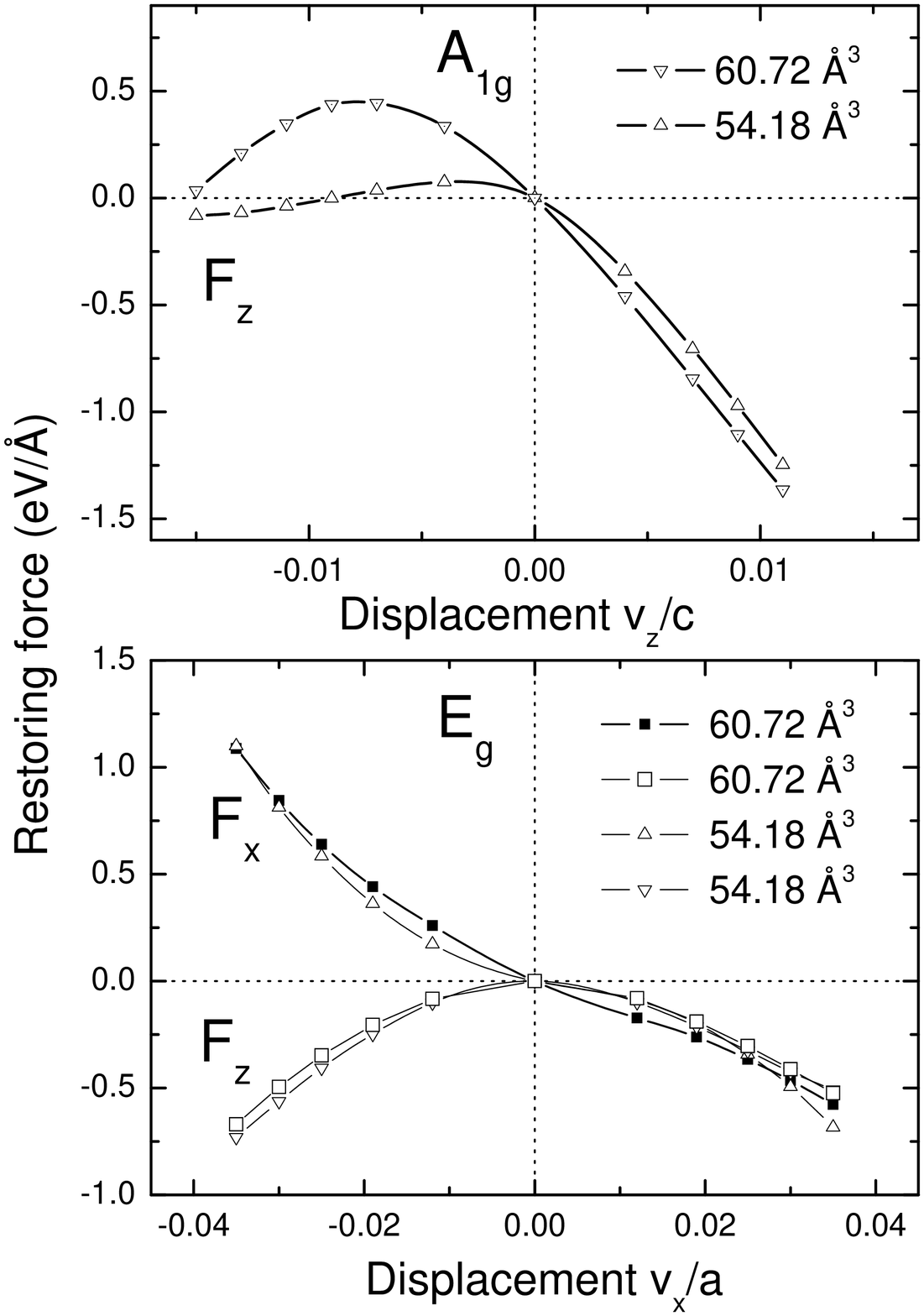}}
\end{minipage}
\caption{\label{fig6} Results of total energy calculations (LDA) for Sb
in the $A7$ structure. Left: Energy at different volumes as a function
of phonon displacement for the A$_{\rm 1g}$ and E$_{\rm g}$ modes. In
the order of decreasing volume, the calculated pressures are -1.68,
-0.19, 1.52, 4.14, and 5.46 GPa. In the A$_{\rm 1g}$ mode, the positive
displacement corresponds to the shortening of the covalent Sb--Sb bond.
Left: Restoring forces as a function of displacement for the A$_{\rm
1g}$ and E$_{\rm g}$ modes at two different volumes. In the case of the
E$_{\rm g}$ mode, both the forces in $x$ and $z$ direction are given
($F_y = 0$ at all volumes). For the A$_{\rm 1g}$ mode, there is no
non-parallel force component.}
\end{figure*}

\section{Phonon frequency calculations}

The phonon frequency calculations presented here are based on an
evaluation of the total energy within the Density Functional Theory
(DFT) \cite{Hohenberg64,Kohn65}. We work in a plane-wave basis and use
the Projector Augmented Waves (PAW) method \cite{Bloch94,Kress99} as
implemented in the VASP codes \cite{VASP}. The (pseudo-) potential
\cite{Bloch94,Kress99} of Sb was available from G. Kresse and J.
Joubert \cite{Kress99}. Scalar relativistic corrections are taken into
account indirectly, through the construction of the
{(pseudo-)}potential. The expansion of wave-functions includes
plane-waves with kinetic energies up to the cutoff energy of $E^{PW} =
215$~eV. As antimony is metallic, we used for the Brillouin zone
sampling a relatively dense uniform $14 \times 14 \times 14$ mesh which
corresponds to 280 k-points in the irreducible wedge \cite{Monkh76}. In
order to deal with the partially occupied orbitals we employed the
method of Methfessel and Paxton \cite{Methf89} (to the first order)
with a smearing parameter $\sigma = 0.1$~eV. Whenever pressures are
quoted in the context of calculations, they were obtained from the DFT
directly using the stress theorem \cite{Nielsen83a}. We have tested the
LDA and GGA versions \cite{Perde92a} of the {(pseudo-)}potential --
essentially we repeated all structural calculations (such as the $E(V)$
and $P(V)$ equations of state) in both LDA and GGA. Neither of the two
approximations turned out to be convincingly superior to the other and
most of the phonon calculations reported below relate to LDA unless
stated otherwise.

The phonon frequencies were calculated in the frozen phonon approach
starting from optimized structural parameters. Our optimized structural
parameters at zero pressure (calculated) are given in Table~\ref{tab2}.
Results of structure optimizations at different volumes are discussed
in the Appendix.

The frozen-phonon approach consists in the evaluation of the total
energy $E^{\rm tot}$ of the crystal with frozen-in atomic
displacements. Small displacements $\vec v(1) = \pm (0,0,v)$ and $\vec
v(2) = \mp (0,0,v)$ in the A$_{1g}$ mode and $\vec v(1) = \pm (v,0,0)$
and $\vec v(2) = \mp (v,0,0)$ in the E$_{g}$ mode were applied to the
two atoms of the basis, and the two values obtained for $\Delta E^{\rm
tot}$ [viz. the $\Delta E^{\rm tot}$(outward) and $\Delta E^{\rm
tot}$(inward)] were averaged. The eigenfrequency is then obtained from
the expression for the energy of a harmonic oscillator
\begin{equation}
\Delta E^{\rm tot} = \frac{1}{2}M \omega^2 \left[\vert \vec
v (1) \vert^2 + \vert \vec v(2) \vert^2\right]~,
\end{equation}
where $M$ is the atomic mass of Sb. As all the $\Delta
E^{tot}(\vec{v})$ variations suggested presence of anharmonic terms we
chose the values of $v_z/c = 0.004$ (for A$_{1g}$) and $v_x/a = 0.012$
(for E$_g$) at which the eventual quartic terms are negligible. The
main anharmonic terms to take care of are the cubic ones, and they are
eliminated by averaging the $\Delta E^{tot}(\rm outward)$ and $\Delta
E^{tot}(\rm inward)$.

Presence of the anharmonicity in the $E(\vec v)$ potential is clearly
visible in Fig.~\ref{fig6}. The scales of $v$ are about the same for
the A$_{1g}$ and E$_g$ mode when measured in absolute displacements
(0.1~\AA \ corresponds, at ambient pressure, to $v_z/c = 0.0089$ or
$v_x/a = 0.023$). We can see that, at low pressures, the energy
variation in the E$_g$ mode does not differ much from a parabolic
shape: the main deviation shows as the slight asymmetry for positive
and negative $v$. The degree of asymmetry remains about the same when
pressure is increased. This behavior is in stark contrast with the one
seen in the A$_{1g}$ mode where, at $P=0$, the $E(\vec v)$ is
essentially symmetrical at small displacements ($v/c \lesssim 0.005$)
but its asymmetry rapidly increases under compression, to reach at $V =
53.14$~\AA$^3$ \ ($P \approx$ 5.5~GPa) an almost flat shape in one of
the displacement directions.

A closer inspection of the A$_{1g}$ potential energy curves reveals, at
low pressures ($P \approx$ 1.58~GPa), a small energy-barrier around
$v_z/c \approx -0.01$; when pressure increases the barrier shifts to
smaller values of $\vert v_z \vert/c$ and becomes lower. An expansion
of the unit cell, on the contrary, makes the barrier higher and shifts
its position to larger values of the displacement (e.g. $v_z =
-0.015\,c \equiv 0.17$~\AA \ at $P = -1.68$~GPa), where it is beyond
the reach of the mean-square displacement at room temperature ($v_z
\approx 0.09$~\AA). It should be noted that the maximum of the barrier
corresponds to the structural parameter $u=0.25$. If we had plotted the
results by measuring the displacement $v_z$ relative to $u=0.25$, we
would have obtained symmetric potential curves similar to those shown
in Refs.\mbox{~}\onlinecite{Chang86} (Sb)
and\mbox{~}\onlinecite{Shick99} (Bi).

More comprehensive information about the anharmonicity is provided by
the variation of \textit{forces} with displacements, which is also
plotted in Fig.~\ref{fig6}. Now the harmonic regime is easily
recognized by the \textit{linear} variation of the restoring force $F$
against $v$. At all volumes close to ambient pressure the A$_{1g}$ mode
is harmonic, until approximately the displacements of $v_z/c = 0.004$,
which is the value we chose for the frozen phonon calculations, to
become conspicuously anharmonic at $V = 54.18$~\AA$^3$ ($P \approx
4.1$~GPa). For the E$_g$ mode, on the other hand, the $F_x$ seems to
deviate little from the linear (i.e. harmonic) behavior for $v_x/a
\lesssim$ 0.02, even when pressure is increased. However, and perhaps
to some surprise at the first sight, a component of force $F_z \ne 0$
appears in this mode.

\begin{figure}[tb]\centering
\includegraphics[width=70mm,clip]{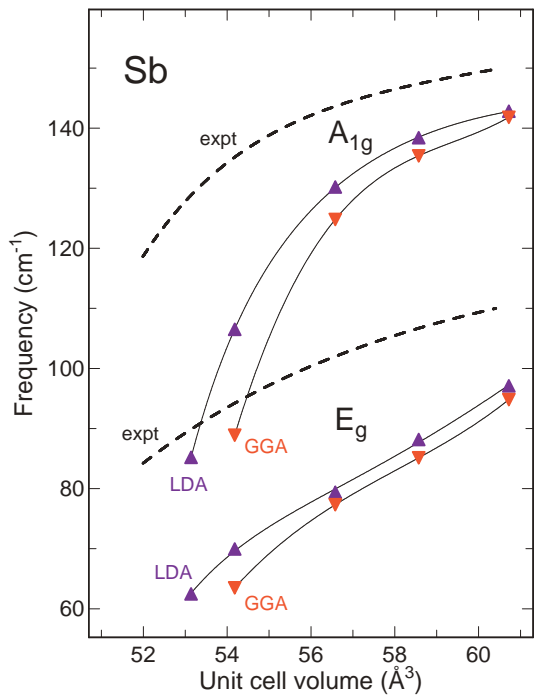}
\caption{\label{fig7} (Color online) Results of frozen-phonon
calculations (LDA and GGA) for Sb in the $A7$ structure: harmonic
phonon frequencies at different unit cell volumes. The experimental
data (dashed lines) are indicated for comparison. Experimental pressure
is converted to volume using a Birch equation of state with
experimental parameters given in Table~\ref{tab2}.}
\end{figure}

This somewhat unexpected emergence of the non-parallel component $F_z$
reminds us that \textit{a multitude} of different anharmonic terms can
appear in the expansion of $E(\vec{v})$, and the presence of
anharmonicity then can show up in two ways: (1) forces are not
\textit{linear} in $v$ and (2) forces are not necessarily
\textit{parallel} to the displacement (see
Ref.\mbox{~}\onlinecite{Kunc85}, Sect. III). It is the ``mixed'' $v_x
v_z$ terms in
\begin{equation}\label{eq:2}\begin{split}
&E^{A_{\rm 1g}}(v_z) = A v_z^2 + B v_z^3 + C v_z^2 v_x + D v_z v_x^2 +
\ldots\\
&\mbox{or}\\
&E^{E_{\rm g}}(v_x) = A v_x^2 + B v_x^3 + C v_x^2 v_z + D v_x v_z^2 +
\ldots
\end{split}\end{equation}
that are responsible for the ``non-parallelism'' of $\vec F$ and $\vec
v$. Figure~\ref{fig6} shows that, while the ``parallel'' term $B$ seems
to be significant in A$_{1g}$, it is the large $C$ which gives rise to
the non-zero component $F_z$ in E$_g$. Dealing only with the
$E(\vec{v})$ variations (Fig.~\ref{fig6}) would, obviously, make us
miss any mixed term in Eqs.~\eqref{eq:2}.

In the harmonic regime, using the Hellmann-Feynman forces merely offers
a convenient, alternative way to explore the calculated $E(\vec{v})$
data. However, in an anharmonic oscillator the knowledge of the
variation $\vec F(\vec{v})$ provides us with additional information on
the mixed terms in Eqs.~\eqref{eq:2}.

In the light of the above discussion, it does not come as a surprise
finding out that the harmonic frequencies calculated from Eq. (1) agree
rather imperfectly with the measured data shown in Fig.~\ref{fig7}.
However, the overall \textit{trend} of the variation is correctly
reproduced. In particular, the initial rate of the variation with
volume is not far from the observed values and the larger nonlinearity
in the softening of A$_{\rm 1g}$ mode is reproduced also. The
discrepancies remind us that the (calculated) harmonic frequencies are
not the quantity that is measured in the spectrum of an anharmonic
oscillator. Another contribution to the discrepancy is caused by a
deviation of the optimized $c/a$ ratios from experimental values (see
Appendix).

An attempt to evaluate the frequency shifts or the line-broadenings in
terms of some of the theories proposed in the past \cite{Cowley68}
would lead us beyond the scope of this paper. In particular, one would
have to assume that anharmonicity is {\it the only} mechanism
responsible for the $\approx$ 20 cm$^{-1}$ shifts appearing in
Fig.~\ref{fig7}, which is by no means certain \cite{Klein82}.

\section{Conclusions}

We have performed a high pressure Raman scattering study of Sb up to
16.6~GPa at room temperature. On the approach to the Sb-I to Sb-II
phase transition near 8 GPa, both the A$_{\rm 1g}$ and E$_{\rm g}$
zone-center phonon modes of the $A7$ phase exhibit a pronounced
softening, of the order -20\%. Frozen-phonon calculations yield
harmonic phonon frequencies in reasonable agreement with the observed
mode softening. The two Raman lines of the A7 phase show an unusually
large broadening under pressure. The fact that the lifetime of phonons
decreases with increasing pressure could become important if, similar
to Bi \cite{Kasami04,Kasami06}, the effect of pressure on coherent
phonon generation in Sb were to become of interest.  We report
Raman-active modes of the phase Sb-II which has a complex guest-host
structure. For the Sb-II phase, all the observed phonon features
exhibit a hardening with increasing pressure, at least up to 16 GPa; a
correlation with the negative pressure coefficient of the
superconducting transition temperature of Sb-II \cite{Witti68} is
apparent.

Perhaps the most interesting question raised by the present study
concerns the coupling of the zone-center optical phonons of the A7
phase of Sb to other phonons (e.g. anharmonic decay) and possibly
electronic processes (electron-phonon coupling). It is the interplay
between mode softening, electronic structure changes at the Fermi
level, and shifts of optical interband transitions (i.e. resonances and
penetration depth) which leads to a more complex scenario of phonon
decay processes compared to semiconducting materials. Our
low-temperature measurements of the Raman line widths indicate that
phonon-related processes are a major cause of the pressure-induced
broadening. So, a first step towards gaining more insight would be to
study the effect of pressure on the phonon dispersion relations,
specifically the acoustic branches.

\appendix*

\section{A7 structure optimization}

The literature about different phases of antimony includes several
studies based on \textit{first-principles} calculations
\cite{Schwa03,Chang86,Seife95,Haeus02}. An interesting problem had been
the question of the existence of a cubic primitive ($cP1$) structure of
Sb at pressures between 6 and 9~GPa. The $cP1$ structure can be viewed
as a special case of $A7$ with $u = 0.25$ and $c/a = \sqrt 6$. First
suggested experimentally in Ref.\mbox{~}\onlinecite{Kolobxx}, the
reality of the $cP1$ phase was {\em not} confirmed in later diffraction
studies of Sb, e.g. Refs.\mbox{~}\onlinecite{Schif81,Schwa03,Degty04}.
Since then the $A7$ $\to$ Sb II/IV transition is believed to be the
first in the sequence of pressure-induced transitions of Sb, and the
'cubic-primitive' subject has occasionally appeared only in theoretical
studies which slowly ``converged'' to elimination, on energetic
grounds, of the $cP1$ phase from the list of stable phases of antimony.

\begin{figure}[tb]\centering
\includegraphics[width=60mm,clip]{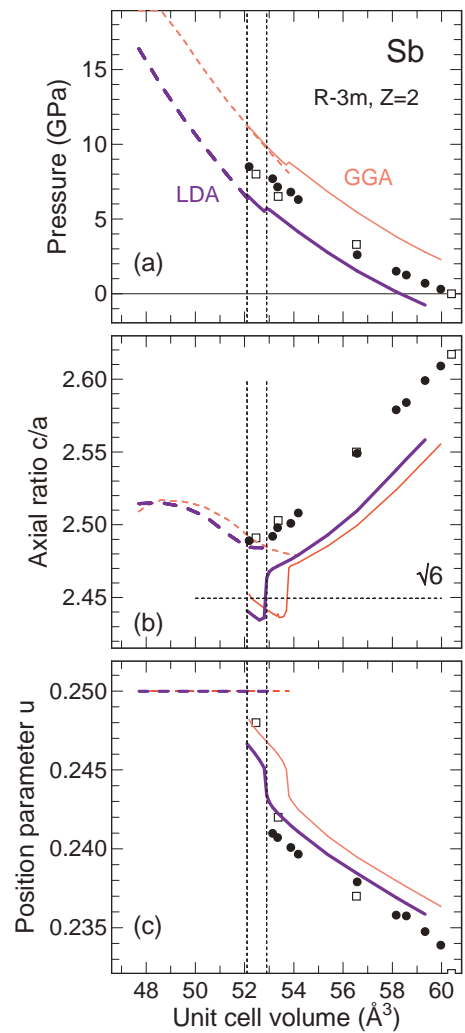}
\caption{\label{fig8} (Color online) Calculated structural properties
of Sb in the A7 structure as a function of rhombohedral unit cell
volume (2 atoms per cell): (a) equation of state $P(V)$; (b) axial
ratio $c/a$; (c) positional parameter $u$. In the LDA approximation, a
sequence of two structural transitions to variants of the $A7$
structure (see text) is predicted between $V = 52.9$ and 52.1 \AA$^3$
(marked by dashed vertical line). In GGA, the corresponding volume
range is larger. Closed circles and open squares represent the
tabulated experimental results \cite{Schif81,Degty04a}. The dashed
vertical line at 52.1 \AA$^3$ coincides with the smallest volume for
which the pure $A7$ phase has been observed experimentally at 300~K
\cite{Schif81}.}
\end{figure}

Our calculated results for the optimized structural properties of Sb-A7
as a function of unit cell volume, in part sampled at very small volume
intervals of 0.1~\AA$^3$, are summarized in Fig.~\ref{fig8}. Let's
first consider the LDA results. Starting from large volume, the
calculated variation $P(V)$ [Fig.~\ref{fig8}(a)] is smooth first. When
passing $V \approx 52.9$~\AA$^3$ a {\em kink} appears in $P(V)$,
followed by narrow interval of smooth variation, until $V =
52.1$~\AA$^3$. The kink in $P(V)$ relates to the evolution of the
structural parameters $c/a$ and $u$ [Figs.~\ref{fig8}(b,c)]. At the
approach to $V = 52.9$ \AA$^3$ the smooth decrease of $c/a$ changes
slope and $c/a$ takes, practically discontinuously, values situated
close to the cubic value $c/a = \sqrt 6$. At the same volume, a nearly
abrupt variation towards the cubic limit $u=0.25$ is also observed in
the behavior of $u$. However, the cubic limit of $u$ is only reached by
a second 'jump' near $V = 52.1$ \AA$^3$. At all volumes $V \le
52.1$~\AA$^3$, the energy-minimization algorithms, set to conserve the
$R\overline{3}m$ symmetry, steer the iteration obstinately towards
exactly $u=0.25$, but with $c/a$ significantly larger again than the
cubic value $c/a = \sqrt 6$. When increasing the volume from below to
above $V = 52.1$ \ \AA$^3$, the optimization procedure tends to stay
trapped at $u=0.25$ until at about $V = 54$~\AA$^3$ it falls back to
the curve obtained for decreasing volume.

The corresponding $E(V)$ calculations (not shown here) tell us that the
range $V \in$ (52.1 ; 52.9)~\AA$^3$ ($P^{\rm LDA}$ between 5.7 and
6.6~GPa) should be considered as the region of co-existence of two $A7$
variants, because the total energy differences are smaller than
2~meV/cell and the specific volumes are almost identical. The third
$A7$ variant is energetically favored for $V\le 52.1$~\AA$^3$ ($P^{\rm
LDA} \ge 6.6$~GPa). However, at these volumes the $A7$ $\to$ Sb-II
transition (not investigated here) is even more favored; the
experiments \cite{Schif81} situate the appearance of the latter phase
at $V = 52.2 \pm 0.02$ \AA$^3$. Hence, the possibility of observing
that third $A7$ variant can be ruled out.

The peculiar variations of $c/a$ and $u$ at $V = 52.9$~\AA$^3$ indicate
a transition to a pseudo-cubic $A7$ variant where only the metric
approaches the cubic value ($c/a = \sqrt 6$). The symmetry of the
structure remains $R\overline{3}m$. The second discontinuity appearing
in $u$ and $c/a$ at $V$ = 52.1 \AA$^3$ suggests a transition towards a
a third A7 variant where the behavior of $c/a$ and $u$ is reversed; the
'cubic' value of $u=0.25$ exactly, constant and independent of volume,
and $c/a > \sqrt{6}$ and varying. So, allowing for the two structural
degrees of freedom of $R\overline{3}m$, our calculations never
converged towards a stable $cP1$ structure of Sb.

The non-smoothness of the $c/a(V)$ and $u(V)$ variations, occurring at
essentially the same two volumes as marked by vertical lines in
Fig.~\ref{fig8}, was spotted in the context of comparisons of LDA $vs.$
GGA calculations reported in Ref.\mbox{~}\onlinecite{Seife95}. The
details of the variations were not pursued throughout the ``instability
region'', and Ref.\mbox{~}\onlinecite{Seife95} concluded at a phase
transition to the $cP1$ structure at $V \le 52$~\AA$^3$ (incorrect),
but found out that ``the axial ratio converges to $c/a \sqrt{6}$ before
the parameter $u$ converges to $u = 1/4$'' (correct). Another, more
recent, calculation \cite{Haeus02} concluded that the $A7$ $\to$ Sb-II
transition takes place at $P = 8.6$~GPa, i.e. {\em before} the
(hypothetical) $A7$ $\to$ $cP1$ transformation might occur at the
(calculated) pressure of $P = 10.7$~GPa. Apparently, non-smooth
variations of the $A7$ structure were not examined in detail in
Ref.\mbox{~}\onlinecite{Haeus02}.

So as to estimate the uncertainties of our above procedures, we briefly
compare the LDA calculations with the ones performed using the GGA.
Within that approximation a similar behavior was obtained but the
``coexistence region'' of $A7$ variants is larger, i.e. $V \in$ (52.2 ;
53.8)~\AA$^3$. A sensible estimate is thus to situate the region of
instabilities, which is to be compared to experiment, as the {\em
average} of the two results: $V \in (52.15 ; 53.35 )$~\AA$^3$. Using
then also for the $V \to P$ conversion the average of the LDA- and
GGA-limits for $P(V)$, we obtain a lower ``critical'' pressure for the
instability region as $P \approx 7.3$ GPa. This is lower than the
experimentally observed appearance of the Sb-II structure (8 to 8.5~GPa
on upstroke). But, taking into account that the Sb-I to Sb-II
transition exhibits a hysteresis of about 1~GPa at room temperature
\cite{Khvos84,Degty04}, the calculated critical pressure is only
slightly less than the equilibrium transition pressure (roughly
estimated as being the average pressure value for the upstroke and
downstroke). So, there is not much room for actually finding an
instability of the A7 phase preceding the Sb-I to Sb-II/IV transition.

In reports on early x-ray diffraction studies performed under
non-hydrostatic conditions \cite{Kolobxx}, the authors uphold to have
observed the sluggish transformation to an intermediate primitive cubic
structure of antimony in the pressure interval $P \in (7.0;
8.5-9.0)$~GPa or at least pronounced anomalies in the evolution of the
$A7$ lattice parameters with pressure. The findings of
Ref.\mbox{~}\onlinecite{Kolobxx}, derived from low-resolution
diffraction diagrams, have not been confirmed in subsequent diffraction
studies. Also, based on discontinuities in the pressure variation of
the electrical resistivity of Sb seen at $P = 7.8$~GPa (upstroke) and
6.7~GPa (downstroke) \cite{Khvos84} the existence of only one
structural transition (A7 to Sb-II) is proposed in the interval
considered here. Nevertheless, the parallel between our calculated
results and the experimental observations of
Ref.\mbox{~}\onlinecite{Kolobxx} is, perhaps, not accidental. It is
conceivable that, in the presence of a substantial uniaxial stress
component along the $C_3$ axis, a precursor transition can be induced.
In this context it should be noted that, at fixed volume, the
calculations yield optimized $c/a$ ratios which are smaller than the
experimental values [cf. Fig.~\ref{fig8}(b)]. So, the condition of zero
deviatoric stresses in the calculations corresponds to a strained
configuration when compared to experimental data. With attention now
being drawn to the importance of uniaxial stress, we arrive at what
could be expected from intuition: the Peierls-like distortion of Sb-I,
being a uniaxial effect, may depend sensitively on uniaxial stress
along $C_3$, with associated structural variations that are not
observed under truly hydrostatic conditions.

To summarize, total energy calculations constrained to space group
$R\overline{3}m$ point to the possible existence of an A7 to A7
transition near the borderline where antimony transforms to the
higher-pressure Sb-II/IV phases. We do not rule out other modes of
instability of the $A7$ structure that require a space group lower in
symmetry (monoclinic)  than $R\overline{3}m$. So, a more general
conclusion would be that the 'normal' $A7$ structure of Sb may exhibit
a structural instability before it transforms to the Sb-II/IV phases.
Some tuning of the phase behavior through uniaxial stress may be
possible. When comparing the calculations with experiment, an inherent
uncertainty concerns the role played by a finite temperature. The
present as well as other \textit{ab initio}
calculations of Sb were performed in the static lattice limit. \\

\begin{acknowledgments}
We thank C. Ulrich for a critical reading of the manuscript. Part of
the computer resources used in this work were provided by the
Scientific Committee of IDRIS (Institut du D\'{e}veloppement et des
Ressources en Informatique Scientifique, Orsay, France).
\end{acknowledgments}


\end{document}